\documentclass[12pt]{article}
\usepackage{amssymb,psfig}

\begin{document}

\title{Stratification of the orbit space in gauge theories. The role of nongeneric
strata}
\author{R.Vilela Mendes \\
{\small Grupo de F\'{\i }sica Matem\'{a}tica}\\
{\small \ Complexo Interdisciplinar, Universidade de Lisboa,}\\
{\small \ Av. Gama Pinto, 2 - P1699 Lisboa Codex, Portugal}\\
{\small e-mail: vilela@cii.fc.ul.pt}}
\date{}
\maketitle

\begin{abstract}
Gauge theory is a theory with constraints and, for that reason, the space of
physical states is not a manifold but a stratified space (orbifold) with
singularities. The classification of strata for smooth (and generalized)
connections is reviewed as well as the formulation of the physical space as
the zero set of a momentum map.

Several important features of nongeneric strata are discussed and new
results are presented suggesting an important role for these strata as
concentrators of the measure in ground state functionals and as a source of
multiple structures in low-lying excitations.

\end{abstract}

\section{Introduction}

There is increasing evidence that gauge theories are Nature's favorite
trick. They have led to a number of questions and some answers of interest
to both physicists and mathematicians. Factorization by local gauge
transformations induces non-trivial bundle structures in gauge theory and,
rather than being a smooth manifold, the gauge orbit space is a stratified
space. It has an open dense generic stratum and several nongeneric strata.
The generic stratum was extensively studied and led to a geometrical
understanding of the Gribov ambiguity\cite{Singer} \cite{Fleischhack1}, the
Faddeev-Popov technique\cite{Babelon} and anomalies\cite{Atiyah}. In
contrast, the role of nongeneric strata has not yet been fully clarified
(see however \cite{Asorey} \cite{Emmrich} \cite{Gaeta2} \cite{Rudolph1}).

In a field theory, physical states are quantum fluctuations around classical
solutions and physical processes are path integrals on the space of field
configurations. Therefore, because of its full measure, the (generalized)
generic connections play the main role in the quantum fluctuations and in
the path integral. However, for the classical solutions around which quantum
fluctuations take place, there is no reason why they cannot be taken from
the nongeneric strata. In fact, the perturbative vacuum is as nongeneric as
it could possibly be.

The main concern in this paper is the role that nongeneric strata play in
the construction of physical states. In Sections 2 and 3 some results are
collected concerning the stratification of gauge orbit spaces and the
characterization of the physical space as the zero set of a momentum map.
Most of these results are widely dispersed in the mathematical literature,
sometimes hidden behind considerable formalism. Hence, a short summary, as
presented in these sections, might be useful.

After a general discussion of possible roles for nongeneric strata, evidence
is presented in Section 4 for their role in the structure of the ground
state measure and low-lying excitations in $SU\left( 2\right) $ and $%
SU\left( 3\right) $ gauge theories.

\section{Stratification of the orbit space in gauge theories}

A classical gauge theory consists of four basic objects:

(i) A principal fiber bundle $P\left( M,G\right) $ with structural group $G$
and projection $\pi :P\rightarrow M$, the base space $M$ being an oriented
Riemannian manifold.

(ii) An affine space $\mathcal{C}$ of connections $\omega $ on $P$, modelled
by a vector space $\mathcal{A}$ of 1-forms on $M$ with values on the Lie
algebra $\mathcal{G}$ of $G$.

(iii) The space of differentiable sections of $P$, called the \textit{gauge
group }$\mathcal{W}$

(iv) A $\mathcal{W}-$invariant functional (the Lagrangian) $\mathcal{L}:%
\mathcal{A}\rightarrow \Bbb{R}$

Choosing a reference connection, the affine space of connections on $P$ may
be modelled by a vector space of $\mathcal{G}$-valued 1-forms ($C^{\infty
}\left( \Lambda ^{1}\otimes \mathcal{G}\right) $). Likewise the curvature $F$
is identified with an element of ($C^{\infty }\left( \Lambda ^{2}\otimes 
\mathcal{G}\right) $).

In coordinates one writes 
\[
A=A_{\mu }^{a}dx^{\mu }t_{a}\qquad x\in M\qquad t_{a}\in \mathcal{G} 
\]
and the action of $\gamma =\{g\left( x\right) \}\in \mathcal{W}$ on $%
\mathcal{A}$ is given by 
\begin{equation}
\gamma :A_{\mu }\left( x\right) \rightarrow \left( gA_{\mu }\right) \left(
x\right) =g\left( x\right) A_{\mu }\left( x\right) g^{-1}\left( x\right)
-\left( \partial g\right) \left( x\right) \cdot g^{-1}\left( x\right)
\label{2.4}
\end{equation}

All statements below refer to the case where $G$ is a compact group.

The action of $\mathcal{W}$ on $\mathcal{A}$ leads to a stratification of $%
\mathcal{A}$ corresponding to the classes of equivalent \textit{orbits} $%
\left\{ gA;g\in \mathcal{W}\right\} $. Let $S_{A}$ denote the \textit{%
isotropy (or stabilizer) group} of $A\in \mathcal{A}$%
\begin{equation}
S_{A}=\left\{ \gamma \in \mathcal{W}:\gamma A=A\right\}  \label{2.5}
\end{equation}
The \textit{stratum} $\Sigma \left( A\right) $ of $A$ is the set of
connections having isotropy groups $\mathcal{W}-$conjugated to that of $A$%
\begin{equation}
\Sigma \left( A\right) =\left\{ B\in \mathcal{A}:\exists \gamma \in \mathcal{%
W}:S_{B}=\gamma S_{A}\gamma ^{-1}\right\}  \label{2.6}
\end{equation}
The \textit{configuration space of the gauge theory} is the quotient space $%
\mathcal{A}/\mathcal{W}$ and therefore a stratum is the set of points in $%
\mathcal{A}/\mathcal{W}$ that correspond to orbits with conjugated isotropy
groups.

The stratification of the gauge space when $G$ is a compact group has been
extensively studied\cite{Kondracki1} - \cite{Rudolph}. The stratification is
topologically regular. The map that, to each orbit, assigns the conjugacy
class of its isotropy group is called the \textit{type}. The set of strata
carries a partial ordering of types, $\Sigma _{\tau }\subseteq \Sigma _{\tau
^{^{\prime }}}$ with $\tau \leq \tau ^{\prime }$ if there are
representatives $S_{\tau }$ and $S_{\tau ^{\prime }}$ of the isotropy groups
such that $S_{\tau }\supseteq S_{\tau ^{\prime }}$. The maximal element in
the ordering of types is the class of the center $Z(G)$ of $G$ and the
minimal one is the class of $G$ itself. Furthermore $\cup _{t\geq \tau
}\Sigma _{t}$ is open and $\Sigma _{\tau }$ is open in the relative topology
in $\cup _{t\leq \tau }\Sigma _{t}$.

Most of the stratification results have been obtained in the framework of
Sobolev connections and Hilbert Lie groups. However, for the calculation of
physical quantities in the path integral formulation 
\begin{equation}
\left\langle \phi \right\rangle =\int_{\mathcal{A}/\mathcal{W}}\phi \left(
\xi \right) e^{i\mathcal{L}\left( \xi \right) }d\mu \left( \xi \right)
\label{2.7}
\end{equation}
a measure in $\mathcal{A}/\mathcal{W}$ is required, and no such measure has
been found for Sobolev connections. Therefore it is more convenient to work
in a space of \textit{generalized connections} $\overline{\mathcal{A}}$,
defining parallel transports on piecewise smooth paths as simple
homomorphisms from the paths on $M$ to the group $G$, without a smoothness
assumption\cite{Ashtekar1}. The same applies to the generalized gauge group $%
\overline{\mathcal{W}}$. Then, there is in $\overline{\mathcal{A}}/\overline{%
\mathcal{W}}$ an induced Haar measure, the Ashtekar-Lewandowski measure\cite
{Ashtekar2} - \cite{Ashtekar3}. Sobolev connections are a dense zero measure
subset of the generalized connections\cite{Marolf}. The question remained
however of whether the stratification results derived in the context of
Sobolev connections would apply to generalized connections. This question
was recently settled by Fleischhack\cite{Fleischhack} who, by establishing a
slice theorem for generalized connections, proved that essentially all
existing stratification results carry over to the generalized connections.
In some cases they even have wider generality.

Because the isotropy group of a connection is isomorphic to the centralizer
of its holonomy group\cite{Booss}, the strata are in one-to-one
correspondence with the Howe subgroups of $G$, that is, the subgroups that
are centralizers of some subset in $G$. Given an holonomy group $H_{\tau }$
associated to a connection $A$ of type $\tau $, the stratum of $A$ is
classified by the conjugacy class of the isotropy group $S_{\tau }$, that
is, the centralizer of $H_{\tau }$ 
\begin{equation}
S_{\tau }=Z\left( H_{\tau }\right)  \label{2.8}
\end{equation}
An important role is also played by the centralizer of the centralizer 
\begin{equation}
H_{\tau }^{\prime }=Z\left( Z\left( H_{\tau }\right) \right)  \label{2.9}
\end{equation}
that contains $H_{\tau }$ itself. If $H_{\tau }^{\prime }$ is a proper
subgroup of $G$, the connection $A$ reduces locally to the subbundle $%
P_{\tau }=\left( M,H_{\tau }^{\prime }\right) $. Global reduction depends on
the topology of $M$, but it is always possible if $P$ is a trivial bundle. $%
H_{\tau }^{\prime }$ is the structure group of the \textit{maximal subbundle}
associated to type $\tau $.Therefore the types of strata are also in
correspondence with types of reductions of the connections to subbundles. If 
$S_{\tau }$ is the center of $G$ the connection is called \textit{irreducible%
}, all others are called \textit{reducible}. The stratum of the irreducible
connections is called the \textit{generic stratum}. It is open and dense and
it carries the full Ashtekar-Lewandowski measure.

\section{Constraints and momentum maps}

\subsection{Singularity structure of Yang-Mills solutions. Linear and
quadratic constraints}

The canonical formulation is the more appropriate one to discuss the role of
non-generic strata in gauge theories. This is because it allows a clear
separation between gauge invariance and the role of constraints. It uses the
theory of bifurcations of zero level sets of momentum mappings as developed
by Arms\cite{Arms1} \cite{Arms2} and Arms, Marsden and Moncrief\cite{Arms3}.
The main points of this construction are summarized below (using an explicit
coordinatewise notation).

With $A^{\mu }=A_{a}^{\mu }t_{a}$ ($\left\{ t_{a}\right\} $ a basis for the
Lie algebra), one takes 
\begin{equation}
\begin{array}{llll}
A_{a}^{i} &  &  &  \\ 
E_{a}^{i} & = & F_{a}^{i0}= & \partial ^{i}A_{a}^{0}-\partial
^{0}A_{a}^{i}-f_{bca}A_{b}^{i}A_{c}^{0}
\end{array}
\label{3.1}
\end{equation}
as canonical variables. If $\mathcal{A}$ is the set of vector potentials in $%
M$ ($M$ is a compact spacelike Cauchy surface or the 3-plane $x^{0}=0$ with
appropriate decaying conditions on the fields at infinity) the set $\left(
A,E\right) $ is a phase-space coordinate in the cotangent bundle of $%
\mathcal{A}$. 
\begin{equation}
\left( A,E\right) \in T^{*}\mathcal{A}\equiv \left( C^{\infty }\left(
\Lambda ^{1}\otimes \mathcal{G}\right) ,C^{\infty }\left( \Lambda
^{2}\otimes \mathcal{G}\right) \right)  \label{3.2}
\end{equation}

Write the Yang-Mills first-order action as 
\begin{equation}
I=\frac{2}{g^{2}}\int d^{4}xTr\left\{ \partial ^{0}A\cdot E+\frac{1}{2}%
\left( E^{2}+B^{2}\right) -A^{0}\left( \nabla \cdot E+\left[ A,E\right]
\right) \right\}  \label{3.3}
\end{equation}
with $B_{a}=-\frac{1}{2}\epsilon _{ijk}F_{a}^{jk}$. Then the Hamiltonian is 
\begin{equation}
H=\int d^{3}x\sum_{a}\left( E_{a}^{2}+B_{a}^{2}\right)  \label{3.4}
\end{equation}
and $A^{0}$ being a Lagrange multiplier, the constraint is 
\begin{equation}
\Gamma \left( x\right) =\nabla \cdot E+\left[ A,E\right] =D\cdot E=0
\label{3.5}
\end{equation}
This being a well posed Cauchy problem, to characterize the singularities of
the solutions it suffices to characterize the singularities of the
constraint equations.

With canonical brackets 
\begin{equation}
\left\{ A_{a}^{i}\left( x\right) ,E_{b}^{j}\left( y\right) \right\}
_{x^{0}=y^{0}}=\delta ^{ij}\delta _{ab}\delta ^{3}\left( x-y\right)
\label{3.6}
\end{equation}
one obtains for the infinitesimal gauge transformations 
\begin{eqnarray}
\delta E_{a}^{i}\left( x\right) &=&f_{bca}\delta \alpha _{b}\left( x\right)
E_{c}^{i}\left( x\right)  \label{3.7} \\
&=&\int d^{3}y\left\{ E_{a}^{i}\left( x\right) ,\sum_{b}\delta \alpha
_{b}\left( y\right) \Gamma _{b}\left( y\right) \right\} _{x^{0}=y^{0}} 
\nonumber
\end{eqnarray}
and 
\begin{eqnarray}
\delta A_{a}^{i}\left( x\right) &=&-\partial ^{i}\delta \alpha _{a}\left(
x\right) +f_{bca}\delta \alpha _{b}\left( x\right) A_{c}^{i}\left( x\right)
\label{3.8} \\
&=&\int d^{3}y\left\{ A_{a}^{i}\left( x\right) ,\sum_{b}\delta \alpha
_{b}\left( y\right) \Gamma _{b}\left( y\right) \right\} _{x^{0}=y^{0}} 
\nonumber
\end{eqnarray}
Therefore $\Gamma _{b}\left( x\right) $ behaves as a Hamiltonian function
for the flow corresponding to the group element generated by $t_{b}$. Hence, 
\begin{equation}
\left( A,E\right) \in T^{*}\mathcal{A}\stackrel{J}{\rightarrow }D\cdot E\in
C^{\infty }\left( \Lambda ^{3}\otimes \mathcal{G}\right)  \label{3.9}
\end{equation}
is what is called a \textit{momentum mapping} for the symmetry group. This
mapping will be denoted $J$. Because $C^{\infty }\left( \Lambda ^{3}\otimes 
\mathcal{G}\right) $ is dual to $C^{\infty }\left( \Lambda ^{0}\otimes 
\mathcal{G}^{*}\right) $ this mapping may also be considered as a mapping
from $T^{*}\mathcal{A}$ to $h^{*}=C^{\infty }\left( \Lambda ^{0}\otimes 
\mathcal{G}^{*}\right) $ (the space of smooth sections on the Lie algebra
dual) 
\begin{equation}
\left( A,E\right) \stackrel{J}{\rightarrow }\Gamma _{b}(x)t^{b}  \label{3.10}
\end{equation}
with $\left\{ t^{b}\right\} $ as a basis for the dual Lie algebra $\mathcal{G%
}^{*}$.

The constraint $D\cdot E=0$ means that the set of solutions of Yang-Mills
theory is the zero set of a momentum mapping.

For the characterization of the set of solutions of the constraint equations
an important role is played by the derivative mapping $J^{\prime }:T_{\left(
A,E\right) }\left( T^{*}\mathcal{A}\right) \rightarrow h^{*}$ and its
adjoint $J^{\prime *}:h^{*}\rightarrow T_{\left( A,E\right) }\left( T^{*}%
\mathcal{A}\right) $. Linearizing $A$ and $E$ around a background field $%
\left( \overline{A},\overline{E}\right) $%
\begin{equation}
\begin{array}{lll}
A & = & \overline{A}+a \\ 
E & = & \overline{E}+e
\end{array}
\label{3.11}
\end{equation}
one easily obtains 
\begin{equation}
\begin{array}{lll}
J^{\prime }\left( a,e\right) _{b} & = & \partial _{i}e_{b}^{i}+f_{bca}\left( 
\overline{A}_{c}^{k}e_{a}^{k}+a_{c}^{k}\overline{E}_{a}^{k}\right) \\ 
\left( J^{\prime *}v\right) _{a}^{k} & = & \left( f_{bac}\overline{E}%
_{c}^{k}v_{b}\left( x\right) ,\left( \partial ^{k}\delta _{ba}+f_{bca}%
\overline{A}_{c}^{k}\right) v_{b}\left( x\right) \right)
\end{array}
\label{3.12}
\end{equation}

Using pointwise metrics on $M$ and $\mathcal{G}$ and integration, a
Riemannian structure is defined in $T^{*}\mathcal{A}$%
\begin{equation}
<<\left( a1,e1\right) ,\left( a2,e2\right) >>=\int d^{3}x\left(
a1a2+e1e2\right)  \label{3.13}
\end{equation}
which is related to the symplectic form by the complex structure 
\begin{equation}
\Bbb{J}\left( a1,e1\right) =\left( -e1,a1\right)  \label{3.14}
\end{equation}
\begin{equation}
\omega \left( (a1,e1),(a2,e2)\right) =<<\Bbb{J}\left( a1,e1\right) ,\left(
a2,e2\right) >>  \label{3.15}
\end{equation}

Because $J^{\prime *}$ is elliptic with injective principal symbol (in $%
\mathcal{L}^{2}$), one has the $\mathcal{L}^{2}$ splittings 
\begin{eqnarray}
T_{\left( \overline{A},\overline{E}\right) }\left( T^{*}\mathcal{A}\right)
&=&\textnormal{Ker}J^{\prime }\oplus \textnormal{Im}J^{\prime *}  \label{3.16} \\
&=&\textnormal{Im}\left( \Bbb{J\circ }J^{\prime *}\right) \oplus \textnormal{Ker}\left(
J^{\prime }\circ \Bbb{J}\right)  \nonumber
\end{eqnarray}
\begin{equation}
h^{*}=\textnormal{Ker}J^{\prime *}\oplus \textnormal{Im}J^{\prime }  \label{3.17}
\end{equation}
One denotes by

$P$ = the projection $T^{*}\mathcal{A}\rightarrow $Im$J^{\prime }$

$H$ = the projection $T^{*}\mathcal{A}\rightarrow $Ker$J^{\prime *}$

Elimination of redundant variables (or gauge fixing) corresponds, in
geometrical terms, to the construction of a slice for the action of the
gauge group $\mathcal{W}$. A slice through a point $\left( \overline{A},%
\overline{E}\right) \in T^{*}\mathcal{A}$ is a submanifold $S\subset T^{*}%
\mathcal{A}$ such that

(i) $\gamma \left( \overline{A},\overline{E}\right) =\left( \overline{A},%
\overline{E}\right) \Longrightarrow \gamma S=S\qquad \gamma \in \mathcal{W}$

(ii) $\gamma S\cap S\neq \emptyset \Longrightarrow \gamma \left( \overline{A}%
,\overline{E}\right) =\left( \overline{A},\overline{E}\right) $

(iii) $T^{*}\mathcal{A}$ is locally the product of the slice $S$ and the
orbit of $\left( \overline{A},\overline{E}\right) $

In this setting the following important results have been obtained\cite
{Arms1} \cite{Arms2} \cite{Arms3}:

(1) An orthogonal slice for the group action is 
\begin{equation}
\left( \overline{A},\overline{E}\right) +\textnormal{Ker}\left( J^{\prime }\circ 
\Bbb{J}\right)  \label{3.18}
\end{equation}
The orthogonal complement, Im$\left( \Bbb{J\circ }J^{\prime *}\right) $, is
the tangent space to the orbit at $\left( \overline{A},\overline{E}\right) $.

(2) Denote by $C=\left\{ J=0\right\} $ the solution set of the constraint
equations and 
\begin{equation}
\begin{array}{lll}
C_{P} & = & \left\{ P\circ J=0\right\} \\ 
C_{H} & = & \left\{ H\circ J=0\right\}
\end{array}
\label{3.19}
\end{equation}
with $C=C_{P}\cap C_{H}$

Then, there is a smooth mapping (the Kuranishi transformation) that maps $%
C_{P}$ locally onto $\left( \overline{A},\overline{E}\right) \oplus $Ker$%
J^{\prime }$. (Ker$J^{\prime }$ is the set of solutions of the linearized
constraints, $\partial _{i}e_{b}^{i}+f_{bca}\left( \overline{A}%
_{c}^{k}e_{a}^{k}+a_{c}^{k}\overline{E}_{a}^{k}\right) =0$).

(3) Ker$J^{\prime *}$, that is, $\left( f_{bac}\overline{E}%
_{c}^{k}v_{b}\left( x\right) =0,\left( \partial ^{k}\delta _{ba}+f_{bca}%
\overline{A}_{c}^{k}\right) v_{b}\left( x\right) =0\right) $, is the set of
infinitesimal symmetries of $\left( \overline{A},\overline{E}\right) $. If
Ker$J^{\prime *}=0$, then $C_{P}=C$ and in this case the solution set of the
full constraint equations is a manifold near $\left( \overline{A},\overline{E%
}\right) $ with tangent space Ker$J^{\prime }$. It means that, if the
background has no symmetries, any solution of the linearized equations
approximates to first order a curve of exact solutions.

(4) In case the background has nontrivial symmetries (Ker$J^{\prime *}\neq 0$%
), define a set $QC$ as follows 
\begin{equation}
QC=\left\{ \left( a,e\right) \in T_{\left( \overline{A},\overline{E}\right)
}\left( T^{*}\mathcal{A}\right) :\left( a,e\right) \in \left( \textnormal{Ker}%
\left( J^{\prime }\circ \Bbb{J}\right) \cap \textnormal{Ker}J^{\prime }\right) 
\textnormal{ and }[a\wedge e]=0\right\}  \label{3.20}
\end{equation}
in coordinates 
\begin{equation}
\begin{array}{rrr}
\partial _{i}a_{b}^{i}+f_{bca}\left( \overline{A}_{c}^{k}e_{a}^{k}+a_{c}^{k}%
\overline{E}_{a}^{k}\right) & = & 0 \\ 
\partial _{i}e_{b}^{i}+f_{bca}\left( \overline{A}_{c}^{k}a_{a}^{k}-e_{c}^{k}%
\overline{E}_{a}^{k}\right) & = & 0 \\ 
\lbrack a\wedge e]_{b}=f_{bca}a_{c}^{k}\left( x\right) e_{a}^{k}\left(
x\right) & = & 0
\end{array}
\label{3.21}
\end{equation}
The first condition is the (gauge fixing) condition that restricts the
perturbation to the slice. The second is the linearized constraint and the
third a quadratic constraint condition.

Then, $S$ being the slice, there is a local diffeomorphism (Kuranishi's) of $%
C\cap S$ onto $\left( \overline{A},\overline{E}\right) +QC$. Equivalently,
the non-linear constraint set is locally $C\thickapprox $Orbit$\left( 
\overline{A},\overline{E}\right) \oplus QC$.

It means that, when the background has non-trivial symmetries, there are
solutions of the linearized equations that are not tangent to actual
solutions and a further quadratic constraint must be imposed on the
perturbations.

(5) The term $[a\wedge e]_{b}=f_{bca}a_{c}^{k}\left( x\right)
e_{a}^{k}\left( x\right) $, used above, is the diagonal of the quadratic
form 
\begin{equation}
J^{\prime \prime }\left( (a1,e1),(a2,e2)\right) =[a1\wedge e2]+[a2\wedge e1]
\label{3.22}
\end{equation}
The degeneracy space of this quadratic form characterizes the solutions with
the same symmetries as $\left( \overline{A},\overline{E}\right) $, namely:

The set of solutions with the same symmetries as $\left( \overline{A},%
\overline{E}\right) $ is a manifold with tangent space at $\left( \overline{A%
},\overline{E}\right) $ given by 
\begin{equation}
\left\{ \left( a,e\right) :\left( a,e\right) \in \left( \textnormal{Ker}\left(
J^{\prime }\circ \Bbb{J}\right) \cap \textnormal{Ker}J^{\prime }\right) \textnormal{ and 
}J^{\prime \prime }\left( (a,e),(a1,e1)\right) =0\right\}  \label{3.23}
\end{equation}
for all $\left( a1,e1\right) \in \left( \textnormal{Ker}\left( J^{\prime }\circ 
\Bbb{J}\right) \cap \textnormal{Ker}J^{\prime }\right) $

\begin{center}
\#
\end{center}

The results listed above have some practical consequences. They mean, for
example, that in perturbative calculations around a background $\left( 
\overline{A},\overline{E}\right) $ with non-trivial symmetries, linear
perturbations must be further restricted by a quadratic condition. In
quantum perturbation theory, the quadratic condition becomes an operator
condition and physical perturbations must be annihilated by the
corresponding quadratic operator.

Further consequences and roles for the non-generic backgrounds are explored
in the remainder of the paper.

\subsection{Confinement and the singlet structure of excitations}

The confinement question covers two distinct statements:

(i) All observables are color neutral

(ii) All physical states all color singlets

For fields transforming under a non-Abelian gauge group, once it is assumed
that gauge invariance is an exact symmetry, the first statement is a simple
manifestation of the existence of a non-Abelian superselection rule. This
has been proved long ago by Strocchi\cite{Strocchi}. Let $\{Q_{a}\}$ be the
set of color charges that generates the (global) gauge group $G$ and $%
\mathcal{O}$ a local observable. Computing the commutator between physical
states $\psi $ and $\phi $,

\begin{equation}
<\psi |[Q_{a},\mathcal{O}]|\phi >=<\psi |[\int d^{3}xJ_{a}^{0},\mathcal{O}%
]|\phi >=<\psi |[\int d^{3}x(J_{a}^{0}-\partial _{i}F_{a}^{i0}),\mathcal{O}%
]|\phi >=0  \label{3.24}
\end{equation}
where the second equality follows from locality of $\mathcal{O}$ and the
third from Gauss' law

\begin{equation}
\partial _{i}F_{a}^{i0}=j_{a}^{0}+gf_{abc}A_{ib}F_{c}^{i0}=J_{a}^{0}
\label{3.25}
\end{equation}
acting on physical states. The term $j_{a}^{0}$ denotes the non-gluonic
charge.

Eq.(\ref{3.24}) implies that all local observables, in the physical space,
commute with the color charges, that is, $G$ is a non-Abelian superselection
rule. In particular it implies that (local) color charges cannot be
observable quantities. Therefore the fact that color charges are not
observable is not a dynamical question, in the sense that it does not depend
on the detailed dynamics of non-Abelian gauge theory but simply on the fact
that current conservation occurs in a particular form, namely the current is
the divergence of an antisymmetric tensor. Unobservability of color charges
is therefore a trivial consequence of non-Abelian gauge symmetry. The deep
question is of course why there is an exactly conserved color gauge symmetry.

It is possible that the second of the confinement statements, the existence
of just color singlets, may also be a ``kinematical'' consequence of gauge
symmetry, although the situation here is not so obvious. The existence of a
non-Abelian superselection rule implies that the superselection sectors are
labelled by the eigenvalues of the Casimir operators. For all except the
singlet sector, there will be more than one vector corresponding to the same
physical state. Hence if non-singlet states were to exist, their description
would imply a departure from the usual quantum mechanical framework. Namely
there would not exist a complete commuting set of observables and the
description of scattering experiments, for example, would require special
care because the computed matrix elements would depend on the initial and
final vector representatives chosen among the physically equivalent
multiplet vectors\cite{Vasilev}. A description using direct integral spaces%
\cite{Vilela2} or some other form of averaging over initial and final
physically equivalent vectors would be mandatory to obtain unambiguous
predictions.

If color is an unbroken symmetry, the question of confinement is not whether
any colored states are going to be found, because color charges are
unobservable anyway, but whether the colorless objects one sees are real
singlets or some sort of balanced admixture of hidden color states. It is
here however that non-generic strata play a role. The quadratic condition $%
f_{bca}a_{c}^{k}\left( x\right) e_{a}^{k}\left( x\right) =0$ means exactly
that the (gluonic) charge associated to excitations around a nongeneric
background is zero. Therefore if the background belongs to a non-generic
stratum, the perturbative vacuum for example, the gluonic low lying
excitations around this background must be singlets. Nothing is said, of
course, concerning excitations around generic states or non-gluonic states.

\subsection{Suppression of non-symmetric fluctuations and wave functional
enhancements}

For quantized gravitational fluctuations around a symmetric background
spacetime, it has been found\cite{Moncrief} that the effect of quadratic
constraints is to suppresses transitions to configurations of lower
symmetry. This led some authors\cite{Cobra} \cite{Emmrich} to conjecture
that the amplitude of the Schr\"{o}dinger functional would display
particular enhancements (or suppressions) near the singularities. This was
illustrated by studying finite-dimensional examples of the Schr\"{o}dinger
equation in configuration spaces with conical singularities.

However, for gauge theories, there is not always suppression of fluctuations
to configurations of lower symmetry. The degeneracy space of the quadratic
form $J^{\prime \prime }$ (Eq.\ref{3.22}) characterizes the fluctuations
with the same symmetry as the background $\left( \overline{A},\overline{E}%
\right) $. But, in addition to this manifold with the same symmetry as $%
\left( \overline{A},\overline{E}\right) $, there are other solutions, with
different symmetry, leading to the conical singularity. For an initial
condition in some stratum, it is known that classical solutions remain in
the same stratum\cite{Otto}, but quantum fluctuations will in principle
explore all the solutions compatible with the linear and quadratic
constraints. Therefore, the conjecture of enhancement near conical
singularities may indeed be true, but it does not necessarily follows from
the theory described above.

In the next section, by studying an approximation to the ground state
functional of SU(2) and SU(3) gauge theories, one finds additional
circumstantial evidence for enhancements near particular classes of
non-generic strata.

\section{Nongeneric strata in SU(2) and SU(3) gauge theories}

\subsection{Ground state functionals in gauge theories}

Using an approximation to the ground state functional, it will be found that
some field configurations, corresponding to reducible strata, concentrate
the ground state measure. The approximation to the ground state functional
is based on an expansion of the path integral representations\cite{Vilela1} 
\begin{equation}
\psi _{0}\left( \chi \right) \sim \int \mathcal{D}\chi \left( \tau \right)
\delta \left( \chi \left( 0\right) -\chi \right) e^{\int_{-\infty
}^{0}L_{E}\left( \chi \left( \tau \right) ,\frac{d}{dt}\chi \left( \tau
\right) \right) d\tau }  \label{4.1}
\end{equation}
or 
\begin{equation}
\left| \psi _{0}\left( \chi \right) \right| ^{2}=\frac{1}{N}\int \mathcal{D}%
\chi \left( \tau \right) \delta \left( \chi \left( 0\right) -\chi \right)
e^{\int_{-\infty }^{\infty }L_{E}\left( \chi \left( \tau \right) ,\frac{d}{dt%
}\chi \left( \tau \right) \right) d\tau }  \label{4.2}
\end{equation}
where $\chi $ denotes the finite or infinite-dimensional set of
configuration space variables, and $L_{E}$ is the Euclidean action. Making
the change of variables 
\begin{equation}
\chi \left( \tau \right) =\chi +z\left( \tau \right)  \label{4.3}
\end{equation}
adding a term $z\left( \tau \right) .J\left( \tau \right) $ to the Euclidean
Lagrangian $L_{E}=-c\left( \frac{d}{dt}\chi \right) ^{2}+V\left( \chi
\right) $, separating the terms quadratic or less than quadratic in $z\left(
\tau \right) $ from higher order terms 
\begin{eqnarray}
&&L_{E}\left( \chi \left( \tau \right) ,\frac{d}{dt}\chi \left( \tau \right)
\right) +z\left( \tau \right) .J\left( \tau \right)  \label{4.4} \\
&=&-V\left( \chi \right) -z\left( \tau \right) .\left( -c\frac{\partial ^{2}%
}{\partial \tau ^{2}}+S\left( \chi \right) \right) z\left( \tau \right)
-\left( \Gamma \left( \chi \right) -J\left( \tau \right) \right) .z\left(
\tau \right) +G\left( z\left( \tau \right) \right)  \nonumber
\end{eqnarray}
and computing the Gaussian integrals for the fluctuations around each
configuration $\chi $, the following representation is obtained for $\left|
\psi _{0}\left( \chi \right) \right| ^{2}$%
\begin{equation}
\left| \psi _{0}\left( \chi \right) \right| ^{2}=\left. \frac{e^{\int d\tau
G(\frac{\partial }{\partial J\left( \tau \right) })}e^{\frac{1}{4}\int d\tau
\left( \Gamma \left( \chi \right) -J\left( \tau \right) \right) \frac{1}{-c%
\frac{\partial ^{2}}{\partial \tau ^{2}}+S\left( \chi \right) }\left( \Gamma
\left( \chi \right) -J\left( \tau \right) \right) }e^{2\sqrt{c}L\left( \chi
\right) \sqrt{S\left( \chi \right) }L\left( \chi \right) }}{\sqrt{\det \frac{%
1}{4\sqrt{cS\left( \chi \right) }}}e^{\int d\tau G(\frac{\partial }{\partial
J\left( \tau \right) })}e^{\frac{1}{4}\int d\tau \left( \Gamma \left( \chi
\right) -J\left( \tau \right) \right) \frac{1}{-c\frac{\partial ^{2}}{%
\partial \tau ^{2}}+S\left( \chi \right) }\left( \Gamma \left( \chi \right)
-J\left( \tau \right) \right) }}\right| _{J=0}  \label{4.5}
\end{equation}
where 
\begin{equation}
L=-\frac{i}{4\sqrt{c}}\int d\tau \frac{1}{\sqrt{S\left( \chi \right) }}%
e^{-\left| \tau \right| \sqrt{\frac{S\left( \chi \right) }{c}}}\left( \Gamma
\left( \chi \right) -J\left( \tau \right) \right)  \label{4.6}
\end{equation}
the evaluation of the Gaussian integrals requiring $S\left( \chi \right) >0$%
. In the $J\rightarrow 0$ limit, $L$ reduces to 
\begin{equation}
L_{0}=-\frac{i}{2}\frac{1}{S\left( \chi \right) }\Gamma \left( \chi \right)
\label{4.7}
\end{equation}

Expanding $\exp \left( \int d\tau G(\frac{\partial }{\partial J\left( \tau
\right) })\right) $, successive approximations to the ground state are
obtained. Of particular interest is the leading term 
\begin{equation}
\left| \psi _{0}\left( \chi \right) \right| _{(0)}^{2}=\left( \det \frac{1}{4%
\sqrt{cS\left( \chi \right) }}\right) ^{-\frac{1}{2}}\exp \left\{ -\frac{%
\sqrt{c}}{2}\Gamma \left( \chi \right) \frac{1}{S\left( \chi \right) }\sqrt{%
S\left( \chi \right) }\frac{1}{S\left( \chi \right) }\Gamma \left( \chi
\right) \right\}  \label{4.8}
\end{equation}
which differs from a perturbative estimate in the fact that, around each
point of the wave functional, a different expansion point is chosen, which
is $\chi $ itself. In the functional integral representation of the ground
state, the wave functional is the integrated effect of paths coming from the
infinite past to the point $\chi $ at $t=0$. Because near $\chi $ the
difference $z\left( \tau \right) -\chi $ is small, the leading term will
contain accurate information from all paths in the neighborhood of $\chi $,
and will be inaccurate only regarding non-harmonic contributions to the
paths far away from $\chi $. For many problems, like the quartic or
exponential oscillators one finds that, up to a shift of parameter values,
the leading term is already practically indistinguishable from the exact
ground state.

For non-Abelian gauge fields one uses the Schr\"{o}dinger formulation for
quantum fields\cite{Schr1} \cite{Schr2}. $\mathcal{A}_{i}^{\alpha }\left(
\tau ,x\right) $ is the space-time Euclidean vector potential, $%
A_{i}^{\alpha }\left( x\right) =\mathcal{A}_{i}^{\alpha }\left( 0,x\right) $
the time-zero field, $\mathcal{B}_{i}^{\alpha }\left( \tau ,x\right)
=\epsilon _{ijk}\left( \partial _{j}\mathcal{A}_{k}^{\alpha }-\frac{g}{2}%
f_{\alpha \beta \gamma }\mathcal{A}_{j}^{\beta }\mathcal{A}_{k}^{\gamma
}\right) $ the nonabelian curvature field and $B_{i}^{\alpha }\left(
x\right) =\epsilon _{ijk}\left( \partial _{j}A_{k}^{\alpha }-\frac{g}{2}%
f_{\alpha \beta \gamma }A_{j}^{\beta }A_{k}^{\gamma }\right) $ its time-zero
counterpart. The time-zero fields are the canonical variables and the
chromoelectric fields the conjugate momenta. Making the change of variables 
\begin{equation}
\mathcal{A}_{i}^{\alpha }\left( \tau ,x\right) =A_{i}^{\alpha }\left(
x\right) +\phi _{i}^{\alpha }\left( \tau ,x\right)  \label{4.9}
\end{equation}
the Euclidean Lagrangian is 
\begin{equation}
L_{E}=-\frac{1}{2}\left( \frac{\partial }{\partial \tau }\phi _{i}^{\alpha
}\right) ^{2}-\frac{1}{2}\left( \mathcal{B}_{i}^{\alpha }\right) ^{2}
\label{4.10}
\end{equation}

To construct the ground state approximation according to Eq.(\ref{4.8}),
notice that 
\begin{equation}
\mathcal{B}_{i}^{\alpha }\left( \tau ,x\right) =B_{i}^{\alpha }\left(
x\right) +\epsilon _{ijk}D_{j}\left( A\right) _{\alpha \beta }\phi
_{k}^{\beta }\left( \tau ,x\right) -\epsilon _{ijk}f_{\alpha \beta \gamma
}\phi _{j}^{\beta }\left( \tau ,x\right) \phi _{k}^{\gamma }\left( \tau
,x\right)  \label{4.11}
\end{equation}
with 
\begin{equation}
D_{j}\left( A\right) _{\alpha \beta }=\partial _{j}\delta _{\alpha \beta
}-gf_{\alpha \beta \gamma }A_{j}^{\gamma }\left( x\right)  \label{4.12}
\end{equation}

Using (\ref{4.8}), the leading term for the ground state functional is 
\begin{equation}
\psi _{0}\left( A\right) _{(0)}=\exp \left\{ -\frac{1}{2}\int
d^{3}xB_{k}^{\alpha }\left( A\left( x\right) \right) \left( \frac{1}{\sqrt{%
R\left( A\left( x\right) \right) .R\left( A\left( x\right) \right) }}\right)
_{kk^{\prime }}^{\alpha \alpha ^{\prime }}B_{k^{\prime }}^{\alpha ^{\prime
}}\left( A\right) \right\}  \label{4.13}
\end{equation}
where the following operator has been defined 
\begin{equation}
R\left( A\right) _{nn^{\prime }}^{\alpha \alpha ^{\prime }}=\epsilon
_{nmn^{\prime }}D_{m}\left( A\right) ^{\alpha \alpha ^{\prime }}
\label{4.14}
\end{equation}

\subsection{SU(2)}

If $G=SU(2)$, the isotropy groups and the structure groups of the maximal
subbundles are : 
\begin{equation}
\begin{tabular}{|c|c|c|}
\hline
& $S_{A}$ & $H_{A}^{^{\prime }}$ \\ \hline
1 & $\Bbb{Z}_{2}$ & $SU(2)$ \\ \hline
2 & $U\left( 1\right) $ & $U\left( 1\right) $ \\ \hline
3 & $SU\left( 2\right) $ & $\Bbb{Z}_{2}$ \\ \hline
\end{tabular}
\label{4.15}
\end{equation}
There are three strata. Stratum 1 is the generic stratum. The other two are
reducible strata.

For particular classes of fields, the leading ground state approximation,
described above, may be given a simple explicit form, which allows us to
test the role of the different strata on the construction of low energy
states. Low-energy states are expected to be associated to fields which, at
least locally, are slowly varying. Therefore a natural subclass to be
studied is the one of constant non-abelian fields $A_{i}^{\alpha }\left(
x\right) =A_{i}^{\alpha }$ restricted to a finite space volume $V$. Consider
the matrix 
\begin{equation}
M_{ij}^{(2)}=\sum_{\alpha =1}^{3}A_{i}^{\alpha }A_{j}^{\alpha }  \label{4.16}
\end{equation}
Being symmetric, this matrix may be diagonalized by a space rotation. As a
result, without loss of generality, $\left\{ A_{1}^{\alpha },A_{2}^{\alpha
},A_{3}^{\alpha }\right\} $ is a set of orthogonal vectors and the $SU\left(
2\right) $ coordinates may be chosen such that 
\begin{equation}
A_{1}^{\alpha }=\left( a_{1},0,0\right) \qquad A_{2}^{\alpha }=\left(
0,a_{2},0\right) \qquad A_{3}^{\alpha }=\left( 0,0,a_{3}\right)  \label{4.17}
\end{equation}
Then 
\begin{equation}
B_{1}^{\alpha }=-g\left( a_{2}a_{3},0,0\right) \qquad B_{2}^{\alpha
}=-g\left( 0,a_{3}a_{1},0\right) \qquad B_{3}^{\alpha }=-g\left(
0,0,a_{1}a_{2}\right)  \label{4.18}
\end{equation}
Using a standard representation for the fractional powers of positive
operators\cite{Yosida} 
\begin{equation}
\frac{1}{\sqrt{R\left( A\right) .R\left( A\right) }}=\frac{1}{\pi }%
\int_{0}^{\infty }d\lambda \lambda ^{-\frac{1}{2}}\frac{1}{\lambda +R\left(
A\right) .R\left( A\right) }  \label{4.19}
\end{equation}
and computing 
\[
B\left( \lambda +R.R\right) B^{T} 
\]
one obtains 
\begin{equation}
\begin{array}{c}
\psi _{0}\left( A\right) _{(0)}=\exp \{-\frac{Vg}{2\pi }\int_{0}^{\infty
}d\lambda \lambda ^{-\frac{1}{2}}[\left( a_{1},a_{2},a_{3}\right) ^{2}\left(
a_{1}^{2}+a_{2}^{2}+a_{3}^{2}\right) \\ 
+\lambda
(a_{1}^{4}(a_{2}^{2}+a_{3}^{2})+a_{2}^{4}(a_{1}^{2}+a_{3}^{2})+a_{3}^{4}(a_{1}^{2}+a_{2}^{2}))
\\ 
+\lambda ^{2}(a_{2}^{2}a_{3}^{2}+a_{1}^{2}a_{3}^{2}+a_{1}^{2}a_{2}^{2})] \\ 
\times [4(a_{1}a_{2}a_{3})^{2}+\lambda (\lambda
+a_{1}^{2}+a_{2}^{2}+a_{3}^{2})^{2}]^{-1}\}
\end{array}
\label{4.20}
\end{equation}
This function is peaked at the zeros of the exponent, which only occur when
two of the $a^{\prime }s$ vanish. For example for $a_{1}=0$ the exponent $%
\sigma \left( a_{1},a_{2},a_{3}\right) $ in $\psi _{0}\left( A\right)
_{(0)}=\exp \{-\frac{Vg}{2}\sigma \left( a_{1},a_{2},a_{3}\right) \}$
becomes 
\begin{equation}
\sigma \left( 0,a_{2},a_{3}\right) =\frac{a_{2}^{2}a_{3}^{2}}{\sqrt{%
a_{2}^{2}+a_{3}^{2}}}  \label{4.21}
\end{equation}

\begin{figure}[htb]
\begin{center}
\psfig{figure=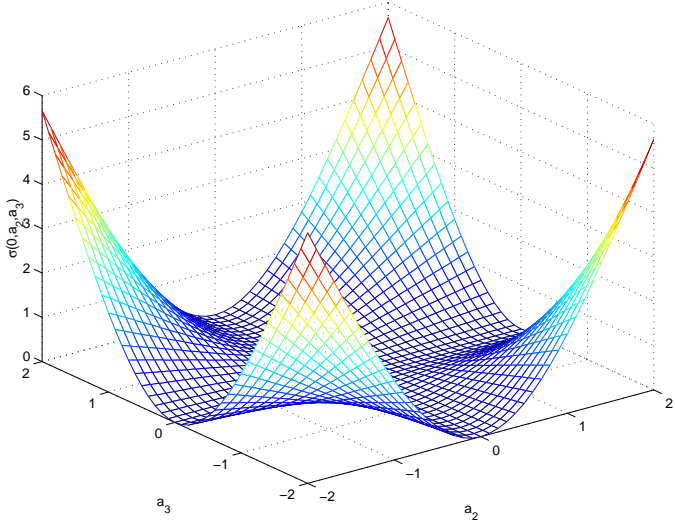,width=9truecm}
\end{center}
\caption{The function $\sigma \left( 0,a_{2},a_{3}\right) 
= \frac{a_{2}^{2}a_{3}^{2}}{\sqrt{a_{2}^{2}+a_{3}^{2}}}$  }
\end{figure}

the function depicted in Fig.1. As a consequence, for this class of fields,
the ground state functional is peaked both near strata of type $2$ and $3$.
Three remarks are in order at this point:

(i) When two of the constants vanish (for example $a_{1}=a_{2}=0$), the
chromomagnetic fields in (\ref{4.18}) also vanish. Therefore a point
strictly on the plane ($a_{1}=a_{2}=0$) might be identified by a gauge
transformation to the origin. However by choosing ($a_{1}=0,a_{2}=%
\varepsilon ,a_{3}=\frac{K}{\varepsilon }$) one obtains, for small $%
\varepsilon $, $\sigma \left( 0,a_{2},a_{3}\right) \simeq \varepsilon K$ and 
$B_{1}^{\alpha }=-g\left( K,0,0\right) $. Therefore arbitrarily large
chromomagnetic fields exist with high probability near the plane ($%
a_{1}=a_{2}=0$). The holonomy group is generated by $\sigma _{1}$ with
centralizer $U\left( 1\right) $. This justifies the statement that there is
a $U\left( 1\right) $ stratum acting as a concentrator of the ground state
measure.

(ii) The same reasoning implies that large field fluctuations are to be
expected in the non-perturbative vacuum and this approximate ground state
functional provides a dynamical view of the vacuum condensates.

(iii) On the other hand, one should not expect all fields in the non-generic
strata to act as concentrators of the ground state measure. A counter
example would be a $U\left( 1\right) $ field with a fast space variation.
Therefore, as stated before, a simple reasoning based on suppression of
transitions by the quadratic constraints cannot be the whole story.

Similar remarks apply to the role played by the nongeneric strata in $%
SU\left( 3\right) $, to be discussed below. That is, in each case, one
identifies the subspace where the ground state functional is peaked and then
finds the corresponding nontrivial neighboring stratum, as in (i) above.

\subsection{SU(3)}

For $G=SU(3)$ the isotropy groups and the structure groups of the maximal
subbundles are\cite{Heil} : 
\begin{equation}
\begin{tabular}{|c|c|c|}
\hline
& $S_{A}$ & $H_{A}^{^{\prime }}$ \\ \hline
1 & $\Bbb{Z}_{3}$ & $SU\left( 3\right) $ \\ \hline
2 & $U\left( 1\right) $ & $U\left( 2\right) $ \\ \hline
3 & $U(1)\times U(1)$ & $U(1)\times U(1)$ \\ \hline
4 & $U\left( 2\right) $ & $U\left( 1\right) $ \\ \hline
5 & $SU\left( 3\right) $ & $\Bbb{Z}_{3}$ \\ \hline
\end{tabular}
\label{4.22}
\end{equation}
There are five strata. Stratum 1 is the generic stratum. All others are
reducible strata. Denote by $\mathcal{Z}$ the gauge group transformations
with values in the center, $\stackrel{\symbol{126}}{\mathcal{W}}=\mathcal{W}/%
\mathcal{Z}$ and $\mathcal{B}$ the space (stratum 1) of irreducible
connections. Then $\mathcal{B}/\stackrel{\symbol{126}}{\mathcal{W}}$ is an
open dense set in $\mathcal{A}/\mathcal{W}$, the complement (the space of
reducible connections) being nowhere dense. Most gauge theory studies
restrict themselves to $\mathcal{B}/\stackrel{\symbol{126}}{\mathcal{W}}$.
However, we will see in a while that, like in SU(2), there are important
contributions from the non-generic strata to low-lying states. In addition
and contrary to the SU(2) case, this structure is not unique, several
possible non-equivalent configurations being possible.

As before one makes a local analysis and considers fields that are constant
in a finite volume $V$. Again, the symmetric matrix 
\begin{equation}
M_{ij}^{(3)}=\sum_{\alpha =1}^{8}A_{i}^{\alpha }A_{j}^{\alpha }  \label{4.23}
\end{equation}
may be diagonalized by a space rotation. $\left\{ A_{1}^{\alpha
},A_{2}^{\alpha },A_{3}^{\alpha }\right\} $ is then a set of three
orthogonal vectors in an eight-dimensional space and there are several
independent choices.\ The stratification of the octet space by SU$\left(
3\right) $ orbits\cite{Michel}, characterizes the independent choices.
However, what is important here is not only SU(3) geometrical independence
but to characterize the choices that lead to qualitatively different ground
state functionals.

(i) Let the only non-zero components be 
\begin{equation}
A_{1}^{1}=a_{1}\qquad A_{2}^{2}=a_{2}\qquad A_{3}^{3}=a_{3}  \label{4.24}
\end{equation}
This case is identical to the one studied for SU(2), being the measure
concentrated now near a stratum with isotropy $U\left( 1\right) \times
U\left( 1\right) $

(ii) Let the non-zero components be 
\begin{equation}
A_{1}^{1}=a_{1}\qquad A_{2}^{2}=a_{2}\qquad A_{3}^{8}=a_{8}  \label{4.25}
\end{equation}
The only non-zero component of the chromomagnetic field is 
\begin{equation}
B_{3}^{3}=-ga_{1}a_{2}  \label{4.26}
\end{equation}
and 
\begin{equation}
B\left( \lambda +R.R\right) B^{T}=\frac{a_{1}^{2}a_{2}^{2}}{\lambda
+a_{1}^{2}+a_{2}^{2}}  \label{4.27}
\end{equation}
In this case, all the fields belong to a $U\left( 1\right) \times U\left(
1\right) $ stratum and the measure being concentrated near $a_{1}=0$ or $%
a_{2}=0$, with the choice $a_{1}=\varepsilon ,a_{2}=\frac{K}{\varepsilon }$
(with $\varepsilon $ small) as before, one proves the existence of $U\left(
1\right) \times U\left( 1\right) $ fields with large measure. This example
shows that not all nongeneric fields are concentrators of the measure. On
the other hand, it seems that whenever the measure is peaked, there is a
nearby nongeneric stratum field.

(iii) If the non-zero components are 
\begin{equation}
A_{1}^{4}=a_{4}\qquad A_{2}^{5}=a_{5}\qquad A_{3}^{8}=a_{8}  \label{4.33}
\end{equation}
\begin{equation}
B_{1}=-g\frac{\sqrt{3}}{2}a_{5}a_{8}\frac{\lambda _{4}}{2}\qquad B_{2}=-g%
\frac{\sqrt{3}}{2}a_{4}a_{8}\frac{\lambda _{5}}{2}\qquad B_{3}=\frac{-g}{2}%
a_{4}a_{5}\left( \frac{\lambda _{3}}{2}+\sqrt{3}\frac{\lambda _{8}}{2}%
\right)   \label{4.33a}
\end{equation}

then

\begin{eqnarray}
&&B\left( \lambda +R.R\right) B^{T}  \label{4.34} \\
&=&4\left\{ 
\begin{array}{c}
\left( 16a_{4}^{2}a_{5}^{2}+12a_{5}^{2}a_{8}^{2}+12a_{4}^{2}a_{8}^{2}\right)
\lambda ^{2} \\ 
+\left(
16a_{4}^{2}a_{5}^{4}+16a_{4}^{4}a_{5}^{2}+12a_{5}^{4}a_{8}^{2}+9a_{5}^{2}a_{8}^{4}+12a_{4}^{4}a_{8}^{2}+9a_{4}^{2}a_{8}^{4}\right) \lambda
\\ 
+12a_{4}^{4}a_{5}^{2}a_{8}^{2}+12a_{4}^{2}a_{5}^{4}a_{8}^{2}+9a_{4}^{2}a_{5}^{2}a_{8}^{4}
\end{array}
\right\} \times  \nonumber \\
&&\left\{ 
\begin{array}{c}
16\lambda ^{3}+\left( 32a_{5}^{2}+32a_{4}^{2}+24a_{8}^{2}\right) \lambda ^{2}
\\ 
+\left(
16a_{4}^{4}+16a_{5}^{4}la+9a_{8}^{4}+32a_{4}^{2}a_{5}^{2}+24a_{8}^{2}a_{4}^{2}+24a_{5}^{2}a_{8}^{2}\right) \lambda +48a_{5}^{2}a_{8}^{2}a_{4}^{2}
\end{array}
\right\} ^{-1}  \nonumber
\end{eqnarray}

and one has the following limits:

For $a_{4}=0$%
\begin{equation}
B\left( \lambda +R.R\right) B^{T}=\frac{12a_{5}^{2}a_{8}^{2}}{4\lambda
+4a_{5}^{2}+3a_{8}^{2}}  \label{4.35}
\end{equation}

For $a_{5}=0$%
\begin{equation}
B\left( \lambda +R.R\right) B^{T}=\frac{12a_{4}^{2}a_{8}^{2}}{4\lambda
+4a_{4}^{2}+3a_{8}^{2}}  \label{4.36}
\end{equation}

For $a_{8}=0$%
\begin{equation}
B\left( \lambda +R.R\right) B^{T}=\frac{4a_{4}^{2}a_{5}^{2}}{\lambda
+a_{4}^{2}+a_{5}^{2}}  \label{4.37}
\end{equation}

For $a_{4}=a_{8}=\varepsilon $ and $a_{5}=\frac{K}{\varepsilon }$ ($%
\varepsilon $ small) the holonomy group is $SU\left( 2\right) $ (V-spin),
the centralizer is $U\left( 1\right) $ (generated by $\sqrt{3}\lambda
_{3}-\lambda _{8}$) and, from (\ref{4.37}), it follows that this $U\left(
1\right) $ field is near a point of high measure. If $a_{8}=0$ with the same
conditions for $a_{4}$ and $a_{5}$, then it would be a field in a $U\left(
1\right) \times U\left( 1\right) $ stratum.

(iv) Finally if 
\begin{equation}
A_{1}=0\qquad A_{2}=a_{2}\left( \frac{\sqrt{2}}{2}\lambda _{4}+\frac{\lambda
_{1}}{2}\right) \qquad A_{3}=a_{3}\left( -\frac{\sqrt{2}}{2}\lambda _{5}+%
\frac{\lambda _{2}}{2}\right)   \label{4.28}
\end{equation}
\begin{equation}
B=g\sqrt{3}a_{2}a_{3}\frac{\lambda _{8}}{2}  \label{4.28a}
\end{equation}
then

\begin{eqnarray}
&&B\left( \lambda +R.R\right) B^{T}  \label{4.29} \\
&=&12\frac{a_{2}^{2}a_{3}^{2}\left( 4\lambda ^{2}+9\lambda \left(
a_{2}^{2}+a_{3}^{2}\right) +18a_{2}^{2}a_{3}^{2}\right) }{\left( 4\lambda
+3a_{2}^{2}+3a_{3}^{2}\right) \left( \lambda ^{2}+3\lambda \left(
a_{2}^{2}+a_{3}^{2}\right) +6a_{2}^{2}a_{3}^{2}\right) }  \nonumber
\end{eqnarray}
All fields in this example belong to a stratum with isotropy group $U\left(
2\right) $. The measure is peaked near $a_{2}=0$ or $a_{3}=0$ and with $%
\left( a_{2}=\varepsilon ,a_{3}=\frac{K}{\varepsilon }\right) $ one finds
nontrivial $U\left( 2\right) $ fields near the maximum of the measure.

In conclusion, one sees that there are geometrical independent choices which
lead to ground state functionals with measures concentrated near each one of
the non-generic strata 2 to 4. Low-lying physical states being represented
by quantum fluctuations around the ground state functional one also
concludes that, at least in the leading order approximation of the expansion
leading to Eq.(\ref{4.13}), there may be distinct classes of excitations
around each type of non-generic strata. This is a much richer structure than
the one implied by the perturbative vacuum (stratum 5). On the other hand
the high probability that is assigned to large chromomagnetic fluctuations
in the ground state functional is consistent with the phenomenological
evidence for the existence of non-trivial vacuum condensates in the QCD
vacuum\cite{Shifman1} \cite{Shifman2}.

\section{Conclusions}

By formulating, in the Hamiltonian formalism, the (primary) constraint as
the zero set of a momentum map, a clear view is obtained of the dual role of
gauge invariance and constraints, as well of the singularity structure of
the stratified orbit space in gauge theories. Some physical consequences of
these results are the need to impose quadratic constraints on perturbation
theory and the natural singlet structure of excitations around non-generic
backgrounds. These are important roles for the nongeneric strata both in
classical and quantum theory. In addition, the role of nongeneric strata on
the structure of anomalies has been discussed in the past\cite{Heil2}.

As a further role for nongeneric strata, there are conjectures concerning
enhancements of the lowest lying Schr\"{o}dinger functional near these
strata. The study of finite-dimensional examples with conical singularities
provided support for this conjecture\cite{Cobra} \cite{Emmrich}. Here, using
a non-perturbative approximation to the ground-state functional in $SU\left(
2\right) $ and $SU\left( 3\right) $ gauge theory, more circumstantial
evidence was provided for this conjecture. In addition to the concentration
of the measure near nongeneric strata, there is also the possibility of a
multiplicity of distinct excitations associated to each stratum type.

Strata of gauge groups $G$ and the structure groups $H$ of subbundles have
been studied in the past in the context of symmetry breaking from to $G$ to $%
H$\hspace{0pt}(see for example Ref.\cite{Isham}). Symmetry breaking
corresponds to a reduction to a subbundle associated to the subgroup. The
possibility of making this reduction depends on the global structure of the
base manifold $M$. This problem is not addressed here, because we have been
concerned mostly with a local analysis. Furthermore in our discussion of
non-trivial vacuum backgrounds, no symmetry breaking is implied. All
equivalent directions in the functional (\ref{4.13}) are equiprobable and
the full gauge symmetry is preserved.

\end{document}